\documentclass[aps,twocolumn]{revtex4}

\usepackage{graphicx}
\usepackage{amssymb}

\begin{document}

\title{Mott State and Quantum Critical Points in Rare-Earth Oxypnictides $RO_{1-x}F_xFeAs$ ($R=La, Sm, Nd, Pr, Ce$)}

\author{Gianluca Giovannetti$^{1,2}$, Sanjeev Kumar$^{1,2}$, Jeroen van den Brink$^{1,3}$}

\address{
$^1$Institute Lorentz for Theoretical Physics, Leiden University, 
          P.O. Box 9506, 2300 RA Leiden, The Netherlands\\ 
$^2$Faculty of Science and Technology and MESA+ Research Institute, University of Twente,
            P.O. Box 217, 7500 AE Enschede, The Netherlands\\ 
$^3$Institute for Molecules and Materials, Radboud Universiteit Nijmegen,
P.O. Box 9010, 6500 GL Nijmegen, The Netherlands}

\date{\today}

\begin{abstract} 
We investigate the magnetic phase diagram of the newly discovered iron-based high temperature oxypnictide superconductors of the type RO$_{1-x}$F$_x $FeAs, with rare earths R=La, Sm, Nd, Pr and Ce by means of {\it ab initio} SGGA and SGGA+U density functional computations. We find undoped LaOFeAs to be a Mott insulator when incorporating electronic correlations via SGGA+U for any physically relevant value of $U$.  The doped compounds are according to SGGA  conductors with a transition from an antiferromagnetic to a non-magnetic state at a hole doping of concentration $x_c$=0.075 for R=Nd, Pr and at electron doping $x_c$=0.25 for Ce and 0.6 for Sm. Superconductivity in these rare-earth oxypnictides thus appears in the vicinity of  a magnetic quantum critical point where electronic correlations are expected to play an important role because of the vicinity of a Mott insulating state at zero doping.
\end{abstract}





\maketitle
{\it Introduction} Very recently, more than two decades after the discovery of  high temperature superconductivity in the cuprates~\cite{bednorz86}, a new class of iron-based superconductors was discovered in the family of rare-earth oxypnictides~\cite{kamihara08,takahashi08}. The highest superconducting temperatures  reported so far are for doped oxypnictides of the type RO$_{1-x}$F$_x$FeAs with  R=Sm, Nd, Pr and Ce  with the transition temperatures of $T_c$=43, 52, 52 and 41 K,  respectively~\cite{ren1,ren2,chen1,chen2}. The Sm compound synthesized under high pressure, was recently reported~\cite{ren3} to have a $T_c$ of 55 K. The lanthanum compound~\cite{kamihara08} has a significantly lower T$_c$ of 26 K.

The electronic structure of these type of oxypnictides was first studied with density functional bandstructure methods by Leb\`egue~\cite{lebegue07}, who investigated the related phosphide LaOFeP, a material with the same crystallographic structure as the iron arsenides.  Even if this compound is build from layers containing formally Fe$^{2+}$ ions,  Leb\`egue showed that this oxypnictide is characterized by large intralayer Fe-P covalency. It was subsequently established that also the electronic bands of the arsenide materials are very similar to the ones in the phosphide~\cite{singh,haule}. The observed large iron $3d$ bandwidth ($\sim$4 eV) strongly reduces the tendency for Fe$^{2+}$ to develop a magnetic moment. In the case of LaOFeP this leads to  the stabilization of a nonmagnetic metallic groundstate~\cite{lebegue07}. This result contrasts with recent first principles calculations on the arsenide compound LaOFeAs. This material was found to be an antiferromagnetic conductor. It was put forward that in the arsenide a spin density wave groundstate forms due to the presence of Fermi surface nesting~\cite{ma,kuroki,dong,mazin}. Here we show that electronic correlations, incorporated within SGGA+U drastically change this picture: for physically relevant values of $U$ LaOFeAs is a magnetic Mott insulator.

The fact that the closely related iron-phosphide and arsenide have different magnetic groundstates, suggests that at zero temperature these materials could be close to an antiferromagnetic - paramagnetic phase transition of the iron 3d moments --a quantum critical point. Our SGGA {\it ab initio} density functional calculations confirm that indeed in the rare-earth oxypnictides with a high superconducting transition temperature, RO$_{1-x}$F$_x$FeAs with  R= Pr, Nd, Sm and Ce, such a quantum critical point is present. In the doped compounds superconductivity appears in the vicinity of this quantum critical point.

\begin{figure}
\centerline{
\includegraphics[width=.65\columnwidth,angle=0]{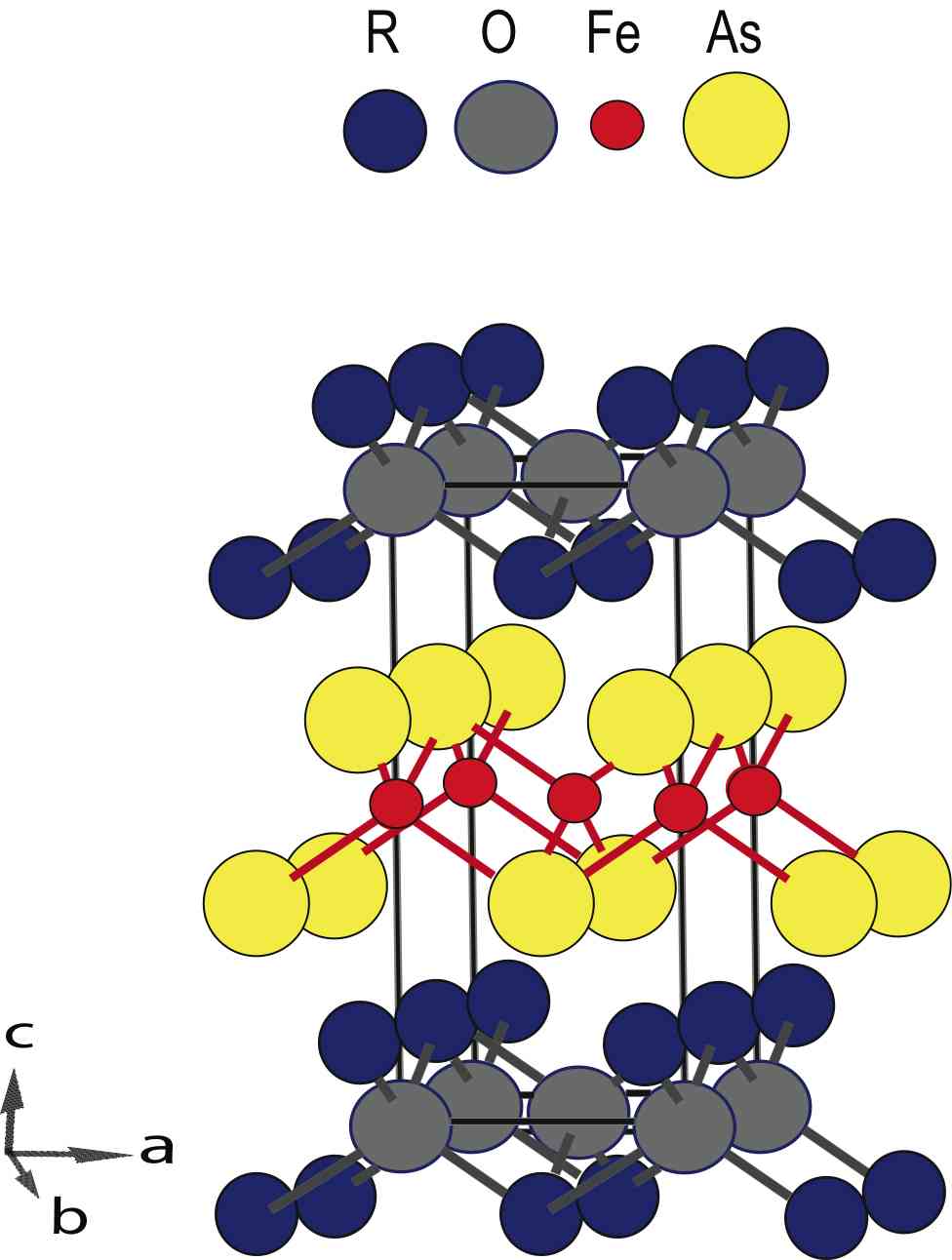}
}
\caption{Schematic view of the crystal structure of ROFeAs  consisting of layers of corner connected FeAs$_4$ tetrahedra (red/yellow) and RO$_4$ tetrahedra (blue/gray).
}
\label{fig:structure}
\end{figure}

\section{Crystal structure} The ROFeAs compounds have a ZrCuSiAs type structure, characterized by Fe-As layers, where Fe is in the center of corner sharing As tetrahedra, see Fig.~\ref{fig:structure}. We compute the relaxed atomic positions, the electronic and the magnetic structure of these systems within the framework of the density functional theory~\cite{HohenberKohn,KohnSham} using the Vienna \textit{ab-initio} simulation package (VASP)~\cite{Kresse}. The Kohn-Sham equations in the self-consistent calculations were solved within a spin generalized gradient approximation (SGGA) for the electronic exchange-correlation potential (PW91 functional~\cite{Perdew}). For the undoped material electronic correlations in the Fe $3d$-shell were accounted for within SGGA+U~\cite{Anisimov,Rohrbach}. The electronic structure is computed using the projector augmented wave method (PAW~\cite{paw1,paw2}), and the valence pseudo-wave-functions are expanded in a plane wave basis set with a cutoff energy of 500 eV. All the integrations in the Brillouin zone are performed with a tetrahedron scheme~\cite{Blochl} using a sampling grid of 12x12x6 k-points and convergency was checked by performing a number of calculations also with a denser grid of k-points. 

\begin{table}
\centerline{
\begin{tabular}{crrrr}
 \hline
   & a (\AA)     & c (\AA)    & $z_{R}$    &  $z_{As}$		\\ \hline
SmOFeAs	&3.940 &    8.496 &    0.1299    	&     0.6402   	\\
NdOFeAs	&3.965 &    8.575 &    0.1427    	&     0.6437   	\\ 
PrOFeAs	&3.985 &    8.595 &    0.1444    	&     0.6417	\\
CeOFeAs	&3.996 &    8.648 &    0.1462    	&     0.6426    	\\
LaOFeAs 	&4.038 &    8.753 &	 0.1415 	&     0.6512     	\\ \hline 
\\
\end{tabular}
}
\label{tab:structure}
\caption{Lattice constants and unit cell parameters of ROFeAs~\cite{quebe00,chen2}, with the experimental $z_{R}$ and $z_{As}$ for La~\cite{kamihara08} and values computed by relaxation otherwise.}
\end{table}

The calculations for ROFeAs (R=La, Sm, Nd, Pr, and Ce) were performed using the experimental lattice parameters~\cite{quebe00}. In cases for which experimental values for the atomic positions within the unit cell are not available, we computed those by minimizing the total energy of the paramagnetic state until the relative changes in energy were less than 10$^{-7}$. The combined results are summarized in Table I. We also computed the relaxed structure in the antiferromagnetic state and within the numerical accuracy we find no differences in atomic positions.  The electronic bandstructures of these systems that we calculated are very similar to the one of LaOFeAs~\cite{singh,haule,ma,kuroki,dong}. Besides the magnetism caused by the iron $3d$ moments that is present in all the FeAs systems that we have studied, there is another contribution to the magnetism from the open $4f$-shells of the rare-earth ions, except for lanthanum. A magnetic interplay between those subsystems can certainly occur, although probably only at very low temperatures as the rare-earth magnetism is generally much weaker than the $3d$ one. Here we wish to focus on the magnetism of the Fe-As subsystem and therefore treat the rare-earth $4f$ electrons in the computations as core electrons, so that they fulfill their role as electron donors but preempting their contribution to magnetism.

\begin{figure}
\centerline{
\includegraphics[width=0.6\columnwidth,angle=0]{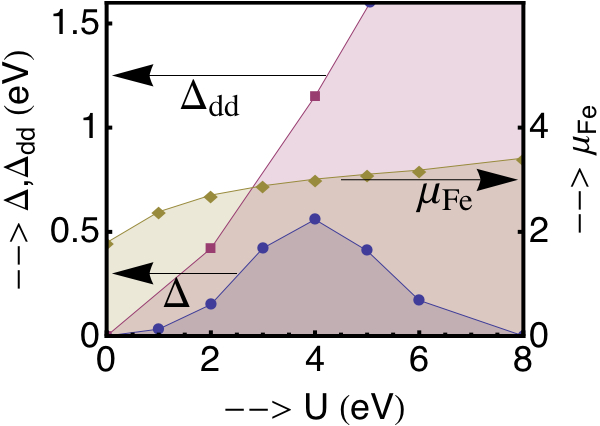}
}
\caption{Gaps $\Delta$ and $\Delta_{dd}$ of LaOFeAs in the magnetic "stripe" phase from SGGA+U as a function of $U$. $\Delta$ is the gap between top of the valence and bottom of the conduction band. For large values of $U$, arsenic bands appear in the gap and closes it. $\Delta_{dd}$ is the splitting between the occupied and empty Fe 3d-bands. Also the local iron moment is shown.}
\label{fig:gaps}
\end{figure}

\section{Electronic and Magnetic Structure of LaOFeAs}
For the undoped ROFeAs compounds we find with SGGA that R=Nd and Pr are nonmagnetic, while Sm, Ce and La are antiferromagnetic, with iron moments of 1.08, 0.66 and 1.62 $\mu_B$, respectively. We also established the validity of the premise of earlier computations that the related iron phosphide LaOFeP is nonmagnetic~\cite{lebegue07}. However, it is well known that in materials containing transition metal ions with open $3d$-shells electronic correlations play an important role. We therefore also compute the electronic structure for LaOFeAs with SGGA+U. In recently studied iron compounds the physical relevant values of $U$ are between 4 and 6 eV; examples are Fe$_3$O$_4$ (5 eV~\cite{leonov04}), Fe$_2$OBO$_3$ (5.5 eV~\cite{leonov05}),  FeO (4.3 eV,~\cite{cococcioni05}) and Fe$_2$SiO$_4$ (4.6 and 4.9 eV,~\cite{cococcioni05}).

\begin{figure}
\centerline{
\includegraphics[width=1.\columnwidth,angle=0]{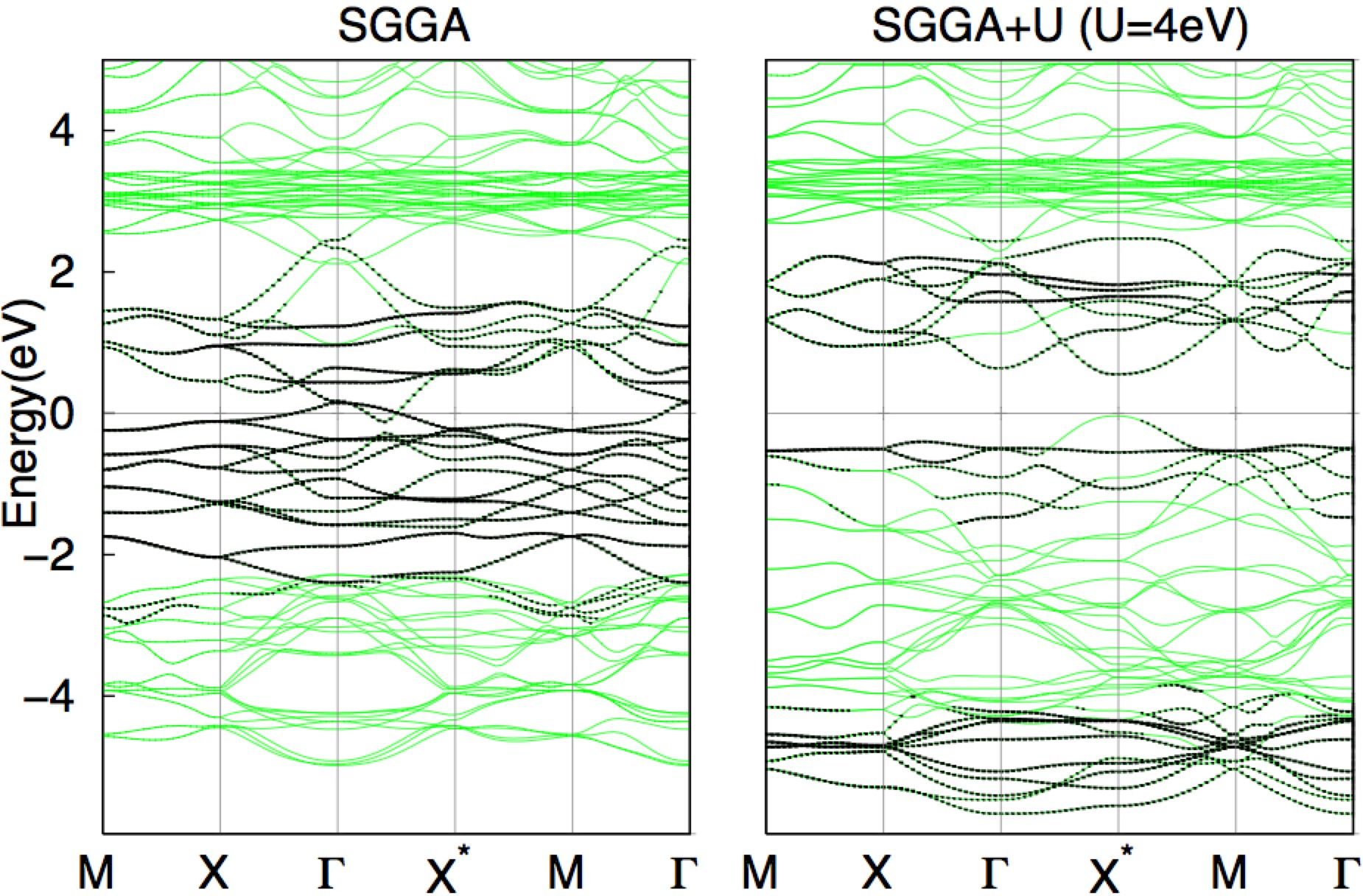}
}
\caption{Bandstructure of undoped LaOFeAs in the magnetic "stripe" phase, computed with SGGA (left) and SGGA+U, using $U=4$eV. Bands with predominatly iron $3d$ character are highlighted in black.}
\label{fig:bands}
\end{figure}

The "striped" magnetic state (with ordering vector $(\pi,0)$ on the squared lattice of iron atoms within the FeAs plane) we find in all cases to be the groundstate, in agreement with experimental observations~\cite{dong,cruz,mcguire}. In this magnetic phase already for very small values of $U$ the system develops a Mott gap, see Fig.~\ref{fig:gaps}. We find for instance for $U$=1 eV that $\Delta$= 33 meV. In the range of $U$ between 4 and 6 eV, the Mott gap is around 400-550 meV.

A remarkable feature is that the gap closes for large values of $U$, around 8 eV.  This contrast with our finding that with increasing $U$ the splitting of the iron $d$-bands increases, as is clear from the value of the splitting between the occupied and empty Fe $3d$-bands $\Delta_{dd}$ shown in Fig.~\ref{fig:gaps}. 
We obtain $\Delta_{dd}$ by computing the dominant character of the bands. From this one can estimate $\Delta_{dd}$, but as bands by definition have a mixed character it is not rigorously defined.  This analysis clearly shows that the reason for the closing of the gap $\Delta$ for large $U$ is that arsenic bands are pushed out of the conduction and valence bands when $U$ increases.  It is evident in the bandstructure and densities of states that these As bands, hybridized with the Fe $3d$ states, start filling up the $d$-$d$ gap.

\begin{figure}
\centerline{
\includegraphics[width=1.\columnwidth,angle=0]{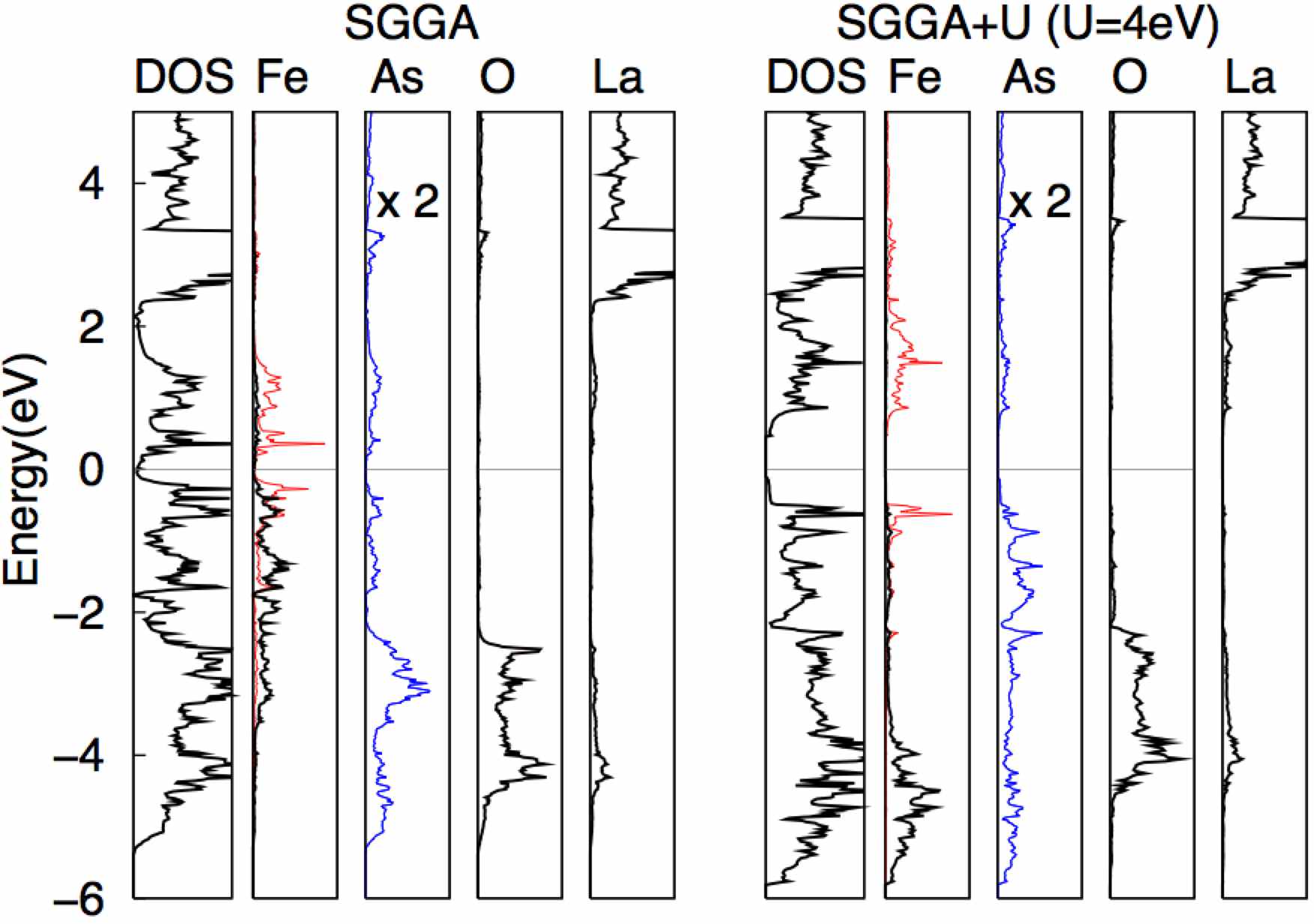}
}
\caption{Density of states (DOS) and projected DOS of undoped LaOFeAs in the magnetic "stripe" phase, computed with SGGA (left) and SGGA+U, using $U=4$eV. For iron the projection along the local Fe magnetic moment is indicated in black, the one antiparallel to it in red.}
\label{fig:dos}
\end{figure}

The bandstructure for $U=0$ and 4 eV is plotted in Fig.~\ref{fig:bands}. The bands and the densities of states (DOS) in Fig.~\ref{fig:dos} reveal that the Hubbard U splits the $3d$ states into three sets of bands: a set of bands (I) below 4eV, a set of bands, (II) at the top of the valence band and (III) at the bottom of the conduction band. We note that there is an anomalously large splitting between sets I and III of around 6 eV, which is considerably larger than the Hubbard $U$ of 4 eV. We will see that this splitting is due to the fact that the local Coulomb interactions have two effects: ($i$) they drive a Mott metal-insulator transition and ($ii$) for small $U$ an increase of the local moment and therefore significantly increase of local spin splittings.  The set I corresponds to quasi-localized 3d orbitals, containing electrons with parallel spins, giving a large contribution to the local Fe moment. The set II corresponds to about one electron per Fe with its spin reversed with respect to the electrons in the set of bands I. Because of the relatively large increase of the local moment due to $U$ the bands I and II become significantly spin split. The unoccupied set of iron bands III is related to empty iron 3d states, with a spin direction that is reversed with respect to  the local moment. These observations are particularly evident from the Fe-DOS plotted in Fig.~\ref{fig:dos}.

For $U$=4 eV around the $X^*$ point a band is pushed out from the valence band that has predominantly arsenic character, mixed to a certain extend with iron $d$. The band has a large dispersion and its contribution to the DOS is small, see Fig.~\ref{fig:dos}. We observe that for larger values of $U$ such a band is pushed out even more until for $U=8$ eV it finally crosses the whole range from valence to conduction bands, spanning the energy scale of $\Delta_{dd}$. The detailed origin of this band and its relevance for the doped systems is an interesting subject for future investigations.

\section{Electronic and Magnetic Structure of Doped Iron-Pnictides}
For doped, metallic systems correlation effects cannot be included in a meanfield like approach such as SGGA+U. For the doped iron-pnictides we therefore only use SGGA computations, which are expected to capture the effects of Fermi surface instabilities.  We find that doping changes the ordered moment markedly, see Fig.~\ref{fig:structure}. We consider both hole and electron doping, which are treated in the computations within the virtual crystal approximation. We find that upon hole doping  the antiferromagnetic state appears in the Nd and Pr compound at a critical concentration $x_c$=0.075. For both the Ce and Sm compound electron doping severely reduces the ordered moment until it vanishes at $x_c$=0.25 and 0.6, respectively, see Fig.~\ref{fig:structure}. The lanthanum system also exhibits a reduction of the ordered moment upon electron doping, although the antiferromagnetic state is stable within the doping range that we consider here. In all cases the ferromagnetic state is higher in energy. 

\begin{figure}
\centerline{\includegraphics[width=.8\columnwidth,angle=0]{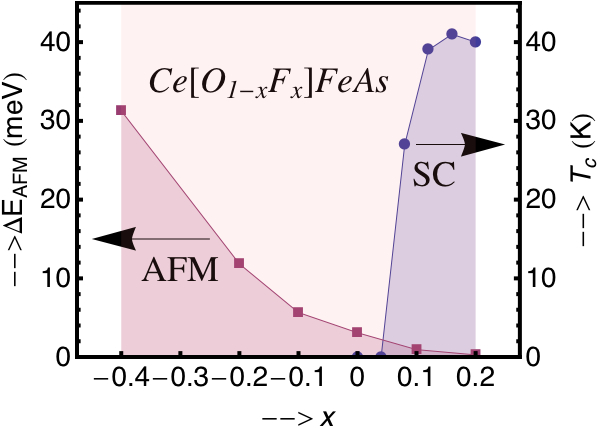}}
\caption{Energy difference per unit cell between nonmagnetic and antiferromagnetic groundstate of doped CeOFeAs, refering to the left vertical scale and the supercoducting transition temperatures~\cite{chen2}, refering to the right scale. The magnetic quantum critical point at $x_c$=0.25 is centered around the superconducting dome.}
\label{fig:energy}
\end{figure}

In Fig.~\ref{fig:energy} the energy difference between the antiferromagnetic and nonmagnetic groundstate is shown together with the experimental doping dependence of the superconducting T$_c$ of CeO$_{1-x}$F$_x$FeAs. Clearly the antiferromagnetic instability extends into the superconducting region, with the magnetic quantum critical point centered in the superconducting dome. In the other compounds the critical point is close to the superconducting region of the phase diagram.

\begin{figure}
\centerline{
\includegraphics[width=.7\columnwidth,angle=0]{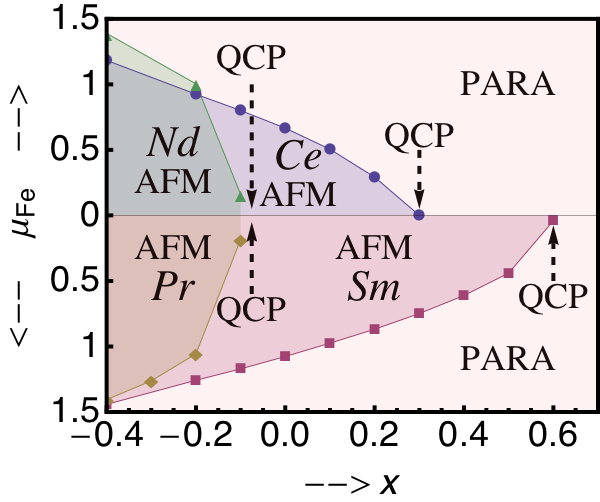}
}
\caption{Ordered iron moment in hole and electron doped ROFeAs, computed from first principles. The antiferromagnetic (AFM) and paramagnetic (PARA) regions are shown for R=Sm, Nd, Pr and Ce. QCP denotes a quantum critical point.
}
\label{fig:structure}
\end{figure}

It has been pointed out by several authors that the antiferromagnetic or spin density wave instability in the  R=La compound is related to the presence of Fermi surface nesting between the Fermi surface regions around the $\Gamma$ and M points in the Brioullin zone~\cite{ma,kuroki,dong,mazin}. The same is found here for R= Sm, Nd, Pr and Ce. The doping dependence of this instability is related to the sensitivity of the nesting condition to the details of crystallographic and electronic structure, which we capture within a single-particle bandstructure approach. 

It should be noted that electron-electron interactions beyond the ones that are effectively taken into account in our SGGA computations, are expected to renormalize the magnitude of the ordered moments and the related critical doping concentrations. Dynamical Mean Field Theory results on doped LaOFeAs for moderate values of $U$ indicate that electronic correlations do affect the hole pockets around the $\Gamma$ and electron pockets around the M point~\cite{haule}. 

Another source of renormalization are the transversal quantum spin fluctuations which one expects to be particularly relevant in these quasi two-dimensional magnetic systems. The treatment of these effects is of course beyond the scope of present day density functional theory and needs to be considered within the framework of model Hamiltonians. One might also expect a small change in $x_c$ when the impurity potentials of the dopants are treated in a scheme beyond the virtual crystal approximation that we have presented here. 

Nevertheless, our calculations make clearly evident that in the high temperature superconducting iron arsenides a magnetic and Mott metal-insulator quantum phase transition is present and that superconductivity appears in the vicinity of those magnetic quantum critical points. The possibility to have a quantum critical state of matter close to a magnetic critical point has recently attracted much attention because the response of such a state is expected to follow universal patterns defined by the quantum mechanical nature of the fluctuations~\cite{sachdev99}. This universality manifests itself through power-law behaviours of response functions, giving rise to scaling laws that are in principle directly accessible experimentally. 

\section{Conclusions}
We find that undoped LaOFeAs is a magnetic Mott insulator with a magnetic "stripe" groundstate for any physically relevant value of the Hubbard $U$ within SGGA+U, suggesting that the doped materials are close to a zero temperature metal-insulator transition. This conclusion is supported by a recent analyses of electronic transport data by Si and Abrahams~\cite{si}.   The Mott gap depends on the precise value of $U$ and is estimated to be between 400 and 550 meV. In addition our SGGA calculations show that doping the rare-earth oxypnictides ROFeAs with R= Sm, Nd, Pr and Ce induces a phase transition from an antiferromagnetic to non-magnetic groundstate.  

It is {\it a priori} not clear whether the magnetic instability competes with the superconducting one and thus suppresses superconductivity or whether it is the root cause of the superconducting state. The fact that it was concluded on the basis of calculations of the electron-phonon coupling strength that a phonon mediated mechanism for superconductivity is unlikely to be operative in these compounds~\cite{boeri,mazin} (although contested by others~\cite{eschrig})  suggests that in these materials the fluctuations associated with the magnetic quantum-critical point are essential for superconductivity. Such a relationship between magnetism, quantum criticality and superconductivity has already been observed in a number of heavy fermion materials, for instance CeRhIn$_5$~\cite{park06}, CeCu$_2$Si$_2$~\cite{gegenwart98,stockert04} and UPd$_2$Al$_3$~\cite{sato01} and uranium intermetallic UCoGe~\cite{huy07}. This suggests a close relation of these materials to the superconducting rare-earth oxypnictides. 

\section{Acknowledgements}
We thank Jan Zaanen, Joel Moore, Frank Kr\"uger, George Sawatzky, Ilya Elfimov and Peter Kes for fruitful discussions. This work was financially supported by {\it NanoNed}, a nanotechnology programme of the Dutch Ministry of Economic Affairs and by the {\it Nederlandse Organisatie voor Wetenschappelijk Onderzoek (NWO)} and the {\it Stichting voor Fundamenteel Onderzoek der Materie (FOM)}. Part of the calculations were performed with a grant of computer time from the {\it Stichting Nationale computerfaciliteiten (NCF)}.

\end{document}